\documentclass{article}
\usepackage{spconf,amsmath,epsfig}

\usepackage{subfigure}
\usepackage{amsfonts}
\usepackage{multirow}
\usepackage{enumerate}
\usepackage{booktabs}
\usepackage{url}

\let\OLDthebibliography\thebibliography
\renewcommand\thebibliography[1]{
  \OLDthebibliography{#1}
  \setlength{\parskip}{0pt}
  \setlength{\itemsep}{0pt plus 0.3ex}
}

\begin{document}\sloppy
\ninept
% Example definitions.
% --------------------
\def\x{{\mathbf x}}
\def\L{{\cal L}}

% Title.
% ------
\title{A Semantic-based Medical Image Fusion}

\name{Fanda Fan\(^{1,2*}\),Yunyou Huang\(^{1,2*}\)\thanks{\(*\)equal contribution.},Lei Wang\(^{1}\),Xingwang Xiong\(^{1,2}\),Zihan Jiang\(^{1,2}\)}{Zhifei Zhang\(^{3^\dagger}\) and Jianfeng Zhan\(^{1,2\dagger}\)\thanks{\(\dagger\)corresponding authors.}}

\address{\(^1\)Institute of Computing Technology Chinese Academy of Sciences, Beijing, China\\
\(^2\)University of Chinese Academy of Sciences, Beijing, China\\
\(^3\)Department of Physiology and Pathophysiology, Capital Medical University, Beijing, China\\
\texttt{\{fanfanda, huangyunyou, zhanjianfeng\}@ict.ac.cn}\\
\texttt{zhifeiz@ccmu.edu.cn}}

\maketitle

\begin{abstract}
  It is necessary for clinicians to comprehensively analyze patient information from different sources.
  Medical image fusion is a promising approach to providing overall information from medical images of different modalities.
  However, existing medical image fusion approaches ignore the semantics of images, making the fused image difficult to understand.
  In this work, we propose a new evaluation index to measure the semantic loss of fused image, and put forward a Fusion W-Net (FW-Net) for multimodal medical image fusion.
  The experimental results are promising: the fused image generated by our approach greatly reduces the semantic information loss, and has better visual effects in contrast to five state-of-art approaches.
  Our approach and tool have great potential to be applied in the clinical setting.
  % We will provide the source code of FW-Net at the Camera Ready stage.
\end{abstract}
\section{Introduction}
\label{sec:intro}

Medical images of different modalities provide different types of information, and they play an increasingly important role in clinical diagnosis.
For example, computed tomography (CT) images display the information of dense structures such as bones and implants, while magnetic resonance (MR) images show high-resolution anatomical information like soft tissue~\cite{yin2018medical}.
Generally, clinicians must thoroughly study medical images of different modalities in order to provide an accurate diagnosis for each patient.
The industry is working towards developing devices with hybrid imaging technologies for obtaining images directly, such as MR/PET and SPECT/CT~\cite{schlemmer2008simultaneous,bockisch2009hybrid}.
However, the devices are not only very expensive, but also difficult to obtain mixed medical images of any two modalities.
Fortunately, there is another low-cost alternative: for each patient, we can fuse existing medical images of different modalities, i.e., CT and MR-T2 (T2 weighted) images.
This alternative is easy to popularize, since it is able to fuse any modal medical image with minimal loss of information~\cite{james2014medical}.

\begin{figure}[t]
  \centering
  \begin{minipage}[t]{1\linewidth}
  \centering
    \subfigure[Source CT image]{
    \includegraphics[width=2.6cm]{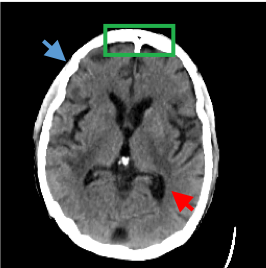}
    }
    \subfigure[Source MR-T2 image]{
    \includegraphics[width=2.6cm]{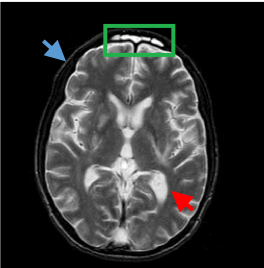}
    }
  \end{minipage}
  \begin{minipage}[t]{1\linewidth}
    \centering
    \subfigure[LP-CNN~\cite{liu2017medical}]{
    \includegraphics[width=2.6cm]{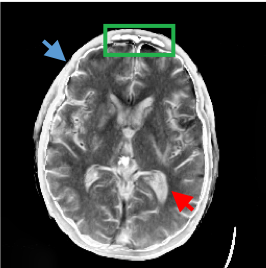}
    }
    \subfigure[NSCT-RPCNN~\cite{das2013neuro}]{
    \includegraphics[width=2.6cm]{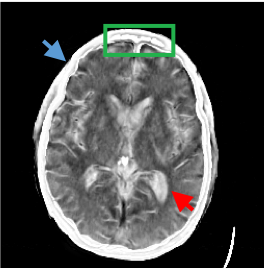}
    }
    \subfigure[NSST-PAPCNN~\cite{yin2018medical}]{
    \includegraphics[width=2.6cm]{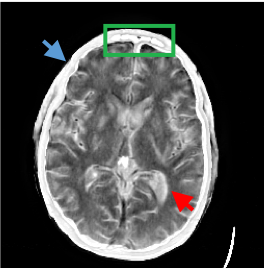}
    }
  \end{minipage}
  \caption{Problems of existing approaches. (a) and (b) are the source images to be fused, and (c) - (e) are the fusion results of the three mainstream approaches.}
  \label{fig:intro_fig}
\end{figure}

In detail, these low-cost approaches are as follows:
On the basis of transformation domain~\cite{liu2017medical,li2013image,bhatnagar2013directive,das2013neuro,yin2018medical},
first, they transform source images into specific coefficients of different scales, and then fuse the coefficients according to several hand-crafted rules, and finally invert the coefficients into fused images.

Unfortunately, there are semantic conflicts in medical images of different modalities, which is overlooked by previous approaches.
The concept of semantics here refers to the fact that brightness in medical images of different modalities represents different meanings.
For example, the brightness of CT image represents the density of tissue, while the brightness of MR-T2 image represents the fluidity and magnetic property of tissue.
So the semantics of brightness in different source images are totally different.
Without resolving the semantic conflicts, the fused images are difficult to read and hence useless in the clinical setting.
Specifically, two significant drawbacks of those approaches are as follows:
(1) The existing approaches overlook semantic conflicts of different source images, which will result in severe semantic loss in the fused images.
In Figure~\ref{fig:intro_fig}, the blue arrow points to a high-density, low-flowing skull, and the red arrow points to a low-density, high-flowing cerebrospinal fluid (in the ventricles).
The semantics of brightness in source images (a) and (b) are significantly different.
However, in the fusion results (c) - (e), there is no difference in the brightness of the skull(blue arrow) and cerebrospinal fluid(red arrow).
(2) The fusion approaches that do not consider semantics of brightness can cause some brain tissue boundaries to blur. %some problem here some
In the green frame of Figure 1 (b), we can clearly see the inflammation area of the frontal sinus, which is the focus of clinicians.
However, since the corresponding part in Figure~\ref{fig:intro_fig} (a) is bright, the frontal sinus boundaries in fusion results (c) - (e) become blurred.

In this paper, motivated by the above issues, we first propose a semantic-based fusion approach.
We provide an autoencoder-based framework, which we name \emph{Fusion W-Net} (FW-Net).
FW-Net encodes all kinds of information extracted from the source image sequence in the fused image as much as possible, and our proposed semantic loss combined with the structural loss in~\cite{ma2015perceptual} can effectively organize the information into a visually fused image.
This work is not the first to combine U-Net~\cite{ronneberger2015u} and autoencoder framework.
W-Net \cite{xia2017w} was proposed for image segmentation task.
However, our FW-Net is different from it in terms of loss function and network structure.
In this paper, we focus on medical image fusion of CT and MR-T2.
However, our approach can be generalized to other images.

Our contributions are as follows:
% (1) We reveal the reason why current image fusion approaches are difficult to apply in the clinical setting, that is, semantic conflicts are ignored.
% (2) We propose a novel FW-Net model to fuse medical images of different modalities.
% (3) A metric is proposed to evaluate the semantic loss in the fused image, and it is used as part of the loss function.
\begin{itemize}
 \item[1)] We reveal the reason why current image fusion approaches are difficult to apply in the clinical setting, that is, semantic conflicts are ignored.
 \item[2)] We propose a novel FW-Net model to fuse medical images of different modalities.
 \item[3)] A metric is proposed to evaluate the semantic loss in the fused image, and it is used as part of the loss function.
\end{itemize}

\section{Proposed Approach}

\begin{figure*}[t]
  \centering
  \centerline{\epsfig{figure=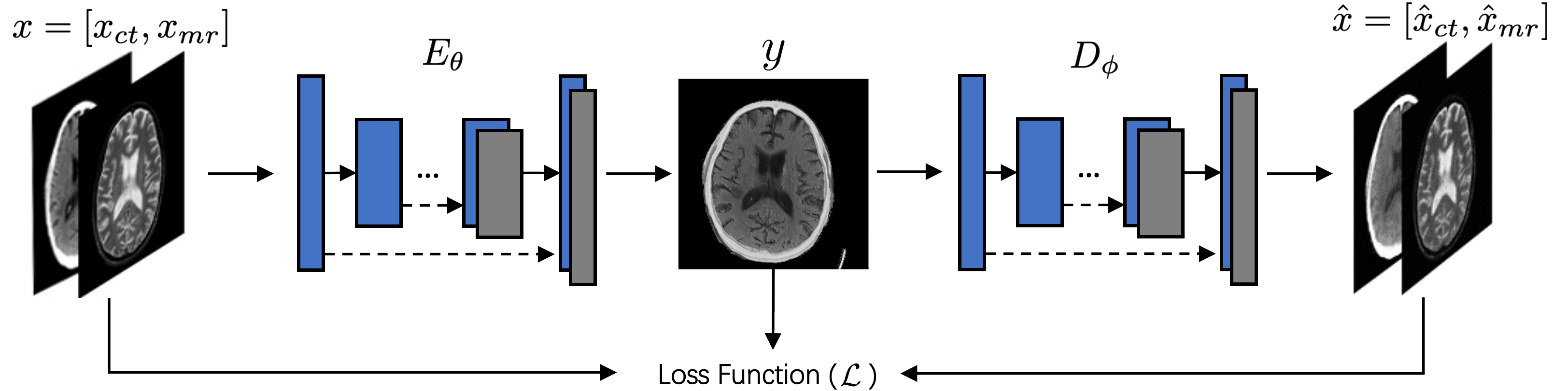,width=14.40cm}}
  \caption{The FW-Net framework.}
  \label{fig:net_fig}
\end{figure*}

The overall framework of the proposed model is shown as Figure~\ref{fig:net_fig}.
A pair of registered images \(x_{ct}\) and \(x_{mr}\) are stacked and fed to the encoder \(E_{\theta}\) in order to generate a fused image \(y\) of the same size as the source images.
And then the decoder \(D_{\phi}\) takes the fused image and generates two reconstructed images \(\hat{x}_{ct}\) and \(\hat{x}_{mr}\).
The framework is an unsupervised end-to-end model.
Our overall objective is
\begin{eqnarray}
  % y &=& ax^2+bx+c \nonumber \\
  % ~ &=& (x+p)(x+q)
  % \mathcal{L}_{total} = \mathcal{L}_{reconstruct}(x_{ct}, x_{mr}, \hat{x}_{ct}, \hat{x}_{mr}) + \nonumber\\
  %  \alpha\mathcal{L}_{MEF\_{SSIM}}(x_{ct}, x_{mr}, y).
  \mathcal{L}_{total} = \mathcal{L}_{reconstruct}(x, \hat{x}) + \alpha \mathcal{L}_{SL}(x, y) + \nonumber\\
  \beta \mathcal{L}_{MEF\_{SSIM}}(x, y), 
\end{eqnarray}
where \(x\) denotes source image sequence, \(\hat{x}\) denotes reconstruct image sequence and \(y\) denotes the fused image.

\(\mathcal{L}_{reconstruct}\) is to measure the difference between the source images and the reconstructed images from pixel-level.
\(\mathcal{L}_{SL}\) and \(\mathcal{L}_{MEF\_{SSIM}}\) measure the semantic and structural loss between source images and the fused image from patch-level.
The reconstruct loss is calculated as 
\begin{eqnarray}
  % y &=& ax^2+bx+c \nonumber \\
  % ~ &=& (x+p)(x+q)
  \mathcal{L}_{reconstruct} = \frac{1}{N} \sum_i (\left\|\hat{x}_{mr}^{(i)} - x_{mr}^{(i)}\right\| + \left\|\hat{x}_{ct}^{(i)} - x_{ct}^{(i)}\right\|),
\end{eqnarray}
where \(\left\| \cdot \right\|\) is the \(\ell_2\) norm of image, \(x^{(i)}\) is the i-th training example, and \(N\) is the number of training samples.
By minimizing the mean squared error (MSE) between the reconstructed images and the source images, the encoder keeps the structure information, texture and more important semantics of the source images to the fused image.
However, the fusion image generated by encoder is not limited to visual considerable.
We add additional \(\mathcal{L}_{SL}\) and \(\mathcal{L}_{MEF\_{SSIM}}\) to constrain the semantics and structure of the fused image, so that the fused image becomes visually impressive.
A ablation study is in Section~\ref{sec:experiments_ablation}.
The following subsection details loss function and the network architecture.

\subsection{Semantic Loss}
\label{sec:semanticloss}
In Section~\ref{sec:intro}, we mentioned that brightness indicates the properties of the tissue in the image.
For example, in CT images, the brightness indicates the density of tissue, while in MR images, brightness indicates the fluidity and magnetic property of tissue.
In the fused image, since the range of pixel values of the fused image and the source image are the same, we need to give more meaning to the brightness in the fused image.
Hence we propose an evaluation index to evaluate the semantic loss \((SL)\) in the fused image.

let \(\{\textbf{x}_k\} = \{\textbf{x}_k | k = 1, 2\}\) denotes the set of image patches extracted from the same spatial location in the source image sequence \(x\), and let \(\textbf{y}\) be the corresponding patch in the fused image \(y\).
Given the source image sequence \(x\) and the fused image \(y\), we define the semantic evaluation index of medical image as follows: 
\begin{eqnarray}
 SL(x, y) = \frac{1}{C} \sum_{i = 1}^M \sum_{j = i + 1}^M \max_{k = 1, 2} ||\mu_{\textbf{x}_k^i} - \mu_{\textbf{x}_k^j}| - |\mu_{\textbf{y}^i} - \mu_{\textbf{y}^j}||, 
\end{eqnarray} % \frac{2}{(M + 1)M} \(\{\textbf{x}_k\}\) and \(\textbf{y}\) are the patches extracted from the same position in \(x\) and \(y\),
where \(M\) is the number of patches in an image, \(\textbf{x}_k^i\) and \(\textbf{y}^i\) are the i-th patch in \(x\) and \(y\), and \(\mu_{\textbf{x}_k^i}\) and \(\mu_{\textbf{y}^i}\) are the mean value of the \(\textbf{x}_k^i\) and \(\textbf{y}^i\) respectively.
\(C\) is used to find the mean, and the value is \(\frac{(M + 1)M}{2}\).
Both \(x\) and \(y\) are normalized to the \([0, 1]\) interval.

\(|\mu_{\textbf{x}_k^i} - \mu_{\textbf{x}_k^j}|\) denotes the brightness difference between two patches under one modality, while \(|\mu_{\textbf{y}^i} - \mu_{\textbf{y}^j}|\) denotes the brightness difference of the corresponding patches in the fused image.
The difference between the two terms indicates whether the semantic changes of one modality are consistent in the fused image.
Considering that the semantics of the fused image is the combination of all the source images, we take the maximum semantic difference between the fused patch and the different modality patches to represents the semantic loss of the fused patch.
Our semantic loss enumerates all combinations of patches to get the final result.
It's worth noting that we have removed the background patches from the calculation because they don't contain any semantic information.
A lower \(SL\) indicates a lower semantic loss of the fused image.
Our semantic loss is as follows:
\begin{eqnarray}
  \mathcal{L}_{SL} = \frac{1}{N} \sum_i SL(x^{(i)}, y^{(i)}).
\end{eqnarray}

For multi-channel images, we need to convert them to YCbCr color channel data, and then compare them in luminance channel.
This is due to the fact that the brightness of the luminance channel changes more significantly than other channels.

\subsection{MEF SSIM Loss}

Multi-exposure image fusion (MEF) is considered as an effective quality enhancement technology, which is widely used in electronic products~\cite{reinhard2010high}.
MEF takes a sequence of images with different exposure levels as input, and synthesizes an image with more information~\cite{burt1984pyramid,burt1993enhanced}.
MEF structural similarity index (MEF SSIM) is proposed by~\cite{ma2015perceptual} to evaluate different MEF algorithms.
After that, Prabhakar~\cite{prabhakar2017deepfuse} proposed a state-of-the-art model, which takes MEF SSIM as the loss function and is used for MEF tasks.
Inspired by~\cite{prabhakar2017deepfuse}, we use MEF SSIM as a part of the loss function, expecting to retain the structure and clearer part of the source images.

The SSIM~\cite{wang2004image} framework divides patch into three parts: structure (\(\textbf{s}\)), luminance (\(l\)) and contrast (\(c\)).
Decompose a given image patch into three components by
\begin{eqnarray}
  \textbf{x}_k &=& \left\| \textbf{x}_k - \mu_{\textbf{x}_k}\right\| \cdot \frac{\textbf{x}_k - \mu_{\textbf{x}_k}}{\left\| \textbf{x}_k - \mu_{\textbf{x}_k}\right\|} + \mu_{\textbf{x}_k} \nonumber \\
      &=& \left\|\tilde{\textbf{x}}_k\right\| \cdot \frac{\tilde{\textbf{x}}_k}{\left\|\tilde{\textbf{x}}_k\right\|} + \mu_{\textbf{x}_k} \nonumber \\
      &=& c_k \cdot \textbf{s}_k + l_k,
\end{eqnarray}
where \(\left\| \cdot \right\|\) denotes the \(\ell_2\) norm of a patch, \(\mu_{\textbf{x}_k}\) is the mean value of the patch, and \(\tilde{\textbf{x}}_k\) is a mean-removed patch.
Since higher contrast value means better image, the desired contrast of the fused image patch \(\hat{c}\) is calculated from:
\begin{eqnarray}
  \hat{c} = \max \limits_{k = 1, 2} \; c_k = \max \limits_{k = 1, 2} \; \left\|\tilde{\textbf{x}}_k\right\|.
\end{eqnarray}
As for the structure information, considering that the corresponding position structure in source images is different, the desired structure of the fused image patch \(\hat{\textbf{s}}\) is a combination of source image patches, which can be obtained by the following formula:
\begin{eqnarray}
  \bar{\textbf{s}} = \frac{\sum_{k = 1}^2 w(\tilde{\textbf{x}}_k) \textbf{s}_k}{\sum_{k = 1}^2 w(\tilde{\textbf{x}}_k)} \quad \text{and} \quad \hat{\textbf{s}} = \frac{\bar{\textbf{s}}}{\left\|\bar{\textbf{s}}\right\|}, 
\end{eqnarray}
where \(w(\cdot)\) is a weighting function, which assigns weights according to the structural consistency between the input patches.
The structure consistency \(R\) is defined by Ma~\cite{ma2015perceptual} to measure the degree of direction agreement among the set of patches.
Its expression is:
\begin{eqnarray}
  R(\{\tilde{\textbf{x}}_k\}) = \frac{\left\|\sum_{k = 1}^2 \tilde{\textbf{x}}_k\right\|}{\sum_{k = 1}^2 \left\|\tilde{\textbf{x}}_k\right\|}.
\end{eqnarray}
It can be observed that \(0 \leq R \leq 1\), and a larger \(R\) value indicates higher structural consistency between patches.
Then \(w(\tilde{\textbf{x}}_k)\) is calculated by
\begin{eqnarray}
  p = \tan \frac{\pi R}{2} \quad \text{and} \quad w(\tilde{\textbf{x}}_k) = \left\|\tilde{\textbf{x}}_k\right\|^p.
\end{eqnarray}
When the structure consistency of patches is low, the weighting function will distribute their weight equally to them.
On the contrary, when all patches have similar structure, patches with high contrast will take up more weight.

Combine the estimated \(\hat{\textbf{s}}\) and \(\hat{c}\) to produce the desired result patch \(\hat{\textbf{x}} = \hat{c} \cdot \hat{\textbf{s}}\).
The quality score of the fused patch is calculated by SSIM~\cite{wang2004image} framework:
\begin{eqnarray}
  Score(\{\textbf{x}_k\}, \textbf{y}) = \frac{2\sigma_{\hat{\textbf{x}}\textbf{y}} + C}{\sigma_{\hat{\textbf{x}}}^2 + \sigma_{\textbf{y}}^2 + C}, 
\end{eqnarray}
where \(\sigma_{\hat{\textbf{x}}}^2\) and \(\sigma_{\textbf{y}}^2\) denote the variances of \(\hat{\textbf{x}}\) and \(\textbf{y}\) respectively, \(\sigma_{\hat{\textbf{x}}\textbf{y}}\) is the covariance between \(\hat{\textbf{x}}\) and \(\textbf{y}\), and \(C\) is a small positive constant.
The final MEF SSIM loss is calculated by the following formula:
\begin{eqnarray}
  \mathcal{L}_{MEF\_{SSIM}} = \frac{1}{N} \sum_i (1 - \sum_{\substack{\{\textbf{x}_k\} \in x^{(i)} \\ \textbf{y} \in y^{(i)}}} Score(\{\textbf{x}_k\}, \textbf{y}))
\end{eqnarray}

\subsection{Network Architecture}
The basic framework of our encoder \(E_{\theta}\) and decoder \(D_{\phi}\) follows the structure of U-Net~\cite{ronneberger2015u}.
U-Net is a fully convolutional network (FCN)~\cite{long2015fully}, which is used for medical image segmentation.
It copies the feature map of layer \(i\) to the layer \(n - i\), where \(n\) is the total number of layers.
The low-level feature map of the network preserves the fine-grained information of the image, while the high-level feature map retains the higher-level semantic information and the high-frequency portion of the image.
It is beneficial to medical image fusion tasks, so we use U-Net in our encoder and decoder.
The stride of \(3 \times 3\) convolution is 1, and its padding is 1.
So after each convolution operation, the size of feature map does not change.
The structure of decoder is almost identical to that of the encoder, except that the input size is \(1 \times 256 \times 256\) and the output size is \(2 \times 256 \times 256\).

We replace the deconvolution operation with a bilinear interpolation operation. %some problem here {a}
Although deconvolution operation can increase the capacity of the model, it makes the quality of the fused image generated by the encoder poor.
The deconvolution operation produces obvious pepper noise and blur in the fused image, while the bilinear interpolation operation produces clearer and smoother images.

\section{Experiments}

\subsection{Experimental settings}
% \subsubsection{Implementation details}
In our FW-net, \(\alpha\) and \(\beta\) are 0.005 and 1 respectively.
In semantic loss calculation, the patch size is \(5 \times 5\), and its stride size is 3.
In MEF SSIM loss calculation, the patch size is \(7 \times 7\), its stride size is 1, and \(C\) is \(9 \times 10^{-4}\).
The batch size is set to 1.
The optimizer is Adam~\cite{kingma2014adam}, where the learning rate is 0.001.
Our FW-Net is implemented in the pytorch framework and run on the Tesla M40 GPU.

% \subsubsection{Approaches of comparison}
We compare our approach with five mainstream algorithms, including the guided filtering-based (GF) approach~\cite{li2013image}, the fuzzy-adaptive reduced pulse-coupled neural network in non-subsampled contourlet (NSCT) domain (NSCT-RPCNN) approach~\cite{das2013neuro}, the phase congruency and directive contrast in NSCT domain (NSCT-PCDC) approach~\cite{bhatnagar2013directive}, the the convolutional neural network in Laplacian pyramid domain (LP-CNN) approach~\cite{liu2017medical}, and the parameter-adaptive pulse coupled neural network in nonsubsampled shearlet domain (NSST-PAPCNN) approach~\cite{yin2018medical}.
The parameters of all these methods are set to the default values from the provided code.

% \subsubsection{Dateset}
We obtained the medical images of the CT and MR-T2 in~\cite{med.harvard.edu}.
The images come from ten people, each with 13 slices, a total of 130 pairs of images.
All source images have the same \(256 \times 256\) pixels with each pair of CT and MR-T2 images aligned and registered.
We used 91 images of 7 people as training set, 26 images of 2 people as validation set and 13 images of 1 person as test set.

% \subsubsection{Metrics}
To assess the quality of the fused image, we evaluate them by the following five indexes:
$Q_{MI}$ \cite{qu2002information} is an entropy-based evaluation index that measures how much information the fused image retains from source images.
\(Q^{AB/F}\) \cite{petrovic2007subjective} is a gradient-based evaluation index, which measures the degree of preservation of edge information of the source images in the fused image.
\(SSIM\) \cite{wang2004image} is an evaluation index based on structural similarity, which measures the structural similarity between the fused image and the source images.
Semantic loss (\(SL\)) is an evaluation index we propose in previous Section~\ref{sec:semanticloss}.
The evaluation indexes \(Q^{AB/F}\), \(SSIM\) and \(SL\) have window sizes of 16, 11 and 3 respectively, and the stride is 1.
Lower is better for \(SL\), higher is better for the others.

% \paragraph*{Approaches of Comparison}

\begin{figure*}[t]
  \centering
  \begin{minipage}[t]{1\linewidth}
  \centering
    \subfigure[Source CT image]{
    \includegraphics[width=3cm]{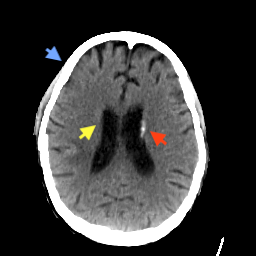}
    }
    \subfigure[Source MR-T2 image]{
    \includegraphics[width=3cm]{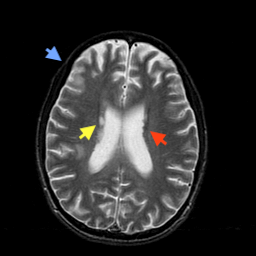}
    }
    \subfigure[LP-CNN]{
    \includegraphics[width=3cm]{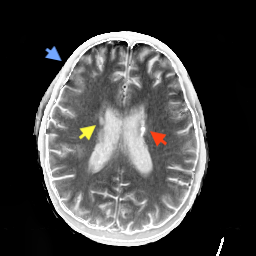}
    }
    \subfigure[NSCT-PCDC]{
    \includegraphics[width=3cm]{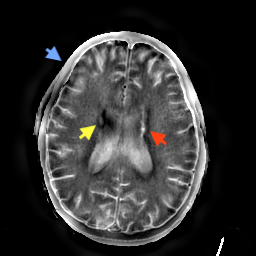}
    }
    \subfigure[NSST-PAPCNN]{
    \includegraphics[width=3cm]{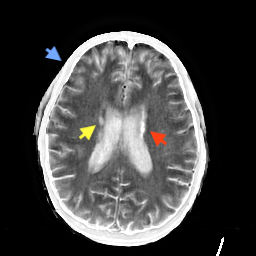}
    }
  \end{minipage}
  \begin{minipage}[t]{1\linewidth}
  \centering
    \subfigure[GF]{
    \includegraphics[width=3cm]{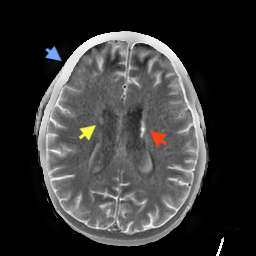}
    }
    \subfigure[NSCT-RPCNN]{
    \includegraphics[width=3cm]{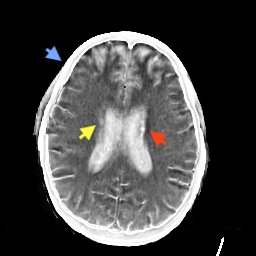}
    }
    \subfigure[FW-Net (Ours)]{
    \includegraphics[width=3cm]{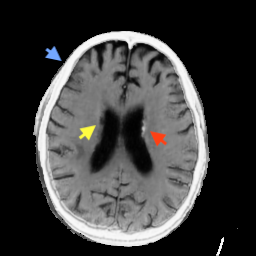}
    }
    \subfigure[FW-Net w/ \(\mathcal{L}_1\)]{
    \includegraphics[width=3cm]{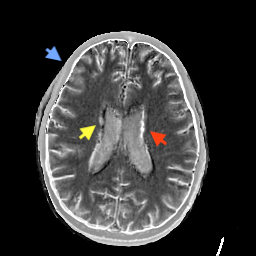}
    }
    \subfigure[FW-Net w/ \(\mathcal{L}_2\)]{
    \includegraphics[width=3cm]{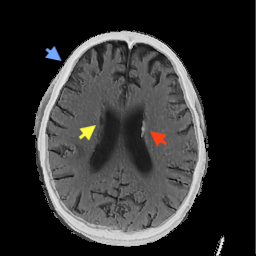}
    }
  \end{minipage}
  \caption{The fused images with different approaches. (a) and (b) are CT and MR-T2 images to be fused, and (c) - (h) are fusion results generated by different approaches.}
  \label{fig:experiments}
\end{figure*}

\subsection{Comparisons with Mainstream Approaches}
\subsubsection{Quantitative evaluation}

\begin{table}
  \caption{Quantitative results of medical image fusion}
  \label{tb:table1}
  \centering
  \begin{tabular}{cccccc}
    \hline
    &                   &        & \multicolumn{2}{c}{\(SSIM\)} \\
    \cline{4-5}
    Approach & \(Q_{MI}\)       & \(Q^{AB/F}\)    & CT              & MR-T2              & SL                  \\
    \hline
    GF       & 0.646           & 0.694          & 0.688          & 0.719              & 0.232            \\
    \hline
    RPCNN    & 0.783           & 0.698          & 0.798          & 0.671              & 0.212            \\
    \hline
    PCDC     & 0.567           & 0.578          & 0.628          & 0.666               & 0.243           \\
    \hline
    LP-CNN   & 0.730           & 0.711          & 0.718          & 0.735               & 0.221       \\
    \hline
    PAPCNN   & 0.731           & 0.631          & 0.743          & 0.719           & 0.217             \\
    \hline
    \textbf{FW-Net} & \textbf{0.853}  & 0.695      & \textbf{0.838} & 0.345   & \textbf{0.134}  \\
    \hline
    \textbf{w/ \(\mathcal{L}_1\)} & 0.719 & \textbf{0.731} & 0.736 & \textbf{0.758}  & 0.229  \\
    \hline
    \textbf{w/ \(\mathcal{L}_2\)} & 0.801  & 0.647 & 0.792 & 0.371 & 0.192  \\
    \hline
    \textbf{w/o \(\mathcal{L}_1\)} & 0.841  & 0.682 & 0.810 & 0.330 &  \textbf{0.134} \\
    \hline
    \textbf{w/o \(\mathcal{L}_2\)} & 0.846  & 0.689 & 0.812 & 0.329 & 0.135  \\
    \hline
    \textbf{w/o \(\mathcal{L}_3\)} & 0.819  & 0.678 & 0.791 & 0.363 & 0.170  \\
    \hline
  \end{tabular}
\end{table}

As the experimental results shown in Table~\ref{tb:table1},  our model has an excellent performance compared with other approaches.
First, there is no doubt that the fused image generated by our approach has the least semantic loss.
Second, our results are the best in \(Q_{MI}\), indicating that our approach retains the information of the source images very well.
Finally, in the \(Q^{AB/F}\) evaluation index, our approach is comparable to other approaches, neither the best nor the worst.
It shows that our approach also preserve the edge information of the source images well. % some problem small
By the way, after training, our approach only needs a forward propagation to get the fused image.
Therefore, as shown in Table~\ref{tb:table2}, our approach has the shortest running time.

Moreover, the \(SSIM\) index shows that compared with other approach, the fused image generated by our approach has the highest structural similarity with the source CT image and the lowest structural similarity with the source MR-T2 image.
This is because the \(SSIM\) index is related to not only the structure information, but also the pixel value.
The fused image generated by our approach tends to represent the new semantic space based on the brightness values of CT.
In the brain structure, most of the brightness values of CT and MR-T2 images are reversed, such as bone and ventricles, which leads to a large gap in scores of \(SSIM\) index.

\subsubsection{Qualitative evaluation}

Figure~\ref{fig:experiments} shows a section of a patient with cerebral toxoplasmosis.
The yellow arrow points to the left ventricle, and the blue arrow points to the outer skull.
It should be noted that the red arrow indicates calcification, which should be the focus of clinicians.
In the source CT image, we can clearly see the bright calcification.
Although this information should be the focus of fused images, the existing approaches tend to retain brightness  and other significant image information regardless of semantics.
Therefore, the bright calcification in CT is mixed with the bright right ventricle in MR-T2, blurring the key information.
It can be seen from the figure that \((c)\), \((d)\), \((e)\) and \((g)\) do not retain the information of calcification well.
Although \((f)\) highlighting the information of calcification, the left and right boundaries of ventricles are blurred.
However, in our approach \((h)\), the brain tissue boundaries and the information of calcification are well preserved. % the calcification information is very good.

Another fundamental issue is semantic conflicts.
For example, the bone (the blue arrow in Figure~\ref{fig:experiments}) appears bright in CT while dark in MR-T2, and the ventricle (the yellow arrow in Figure~\ref{fig:experiments}) appears dark in CT while bright in MR-T2.
The previous approaches do not distinguish the brightness in different source images, which results in the same brightness values of ventricles and skulls in fusion results (c) - (g).
However, they have different semantics in fact.
The semantic conflicts here are reflected in the high-density, low-flowing skull and low-density, high-flowing cerebrospinal fluid (in the ventricles) that present the same brightness values in the fused image.
Our approach resolves the semantic conflicts well, making the brightness of the skull and the cerebrospinal fluid opposite.

\begin{table}
  \caption{Running time of different approach}
  \label{tb:table2}
  \centering
  \begin{tabular}{cccccc}
    \hline
    \cite{li2013image} & \cite{das2013neuro}   & \cite{bhatnagar2013directive}       & \cite{liu2017medical} & \cite{yin2018medical} & FW-Net   \\
    \hline
    0.09       & 12.2           & 28.6          & 21.3          & 9.6              & \textbf{0.01}            \\
    \hline
  \end{tabular}
\end{table}

It is also worth mentioning that our approach produces "cleaner" images, and the brightness is biased towards the source CT image, but the semantics are more abundant.
The thin blood vessels that are not present in the source CT image are well presented in the source MR-T2.
In our fusion result, the thin blood vessels are converted to be dark, indicating their low-density, high-flowing characteristics (the same semantics as cerebrospinal fluid).
Our FW-Net converts the same part of the different modal source images into a new semantic space, eliminating noise caused by inconsistencies in images of different modalities.
This makes fused images shaper and semantically richer.

\subsection{Ablation Study}
\label{sec:experiments_ablation}
In this sub section, we analyze the effect of each loss function.
For simplicity, we use \(\mathcal{L}_1\), \(\mathcal{L}_2\), and \(\mathcal{L}_3\) to represent \(\mathcal{L}_{MEF\_SSIM}\), \(\mathcal{L}_{reconstruct}\), and \(\mathcal{L}_{SL}\) respectively, and use w/ to represent with and w/o to represent without.
It is worth noting that the experiment w/ \(\mathcal{L}_1\) is the same as the DeepFuse~\cite{prabhakar2017deepfuse} approach except that the network structure is different.

% In Table~\ref{tb:table1}, we can see that w/ \(\mathcal{L}_1\) well retains the edge information in the source images, and the \(SSIM\) show a relatively balanced state.
In Table~\ref{tb:table1}, the results show that \(\mathcal{L}_1\) loss keeps the edge and structure information of the source images well, but it also finds that there is almost no semantic information.
It is shown in (i) of Figure~\ref{fig:experiments}.
\(\mathcal{L}_2\) loss has the ability to retain all kinds of information in the source images, but due to the lack of semantic and structural constraints, the performance is not so well.
It is shown in (j) of Figure~\ref{fig:experiments}.
In addition to retaining the semantics in the source images, \(\mathcal{L}_3\) loss also prevents the fused image boundary from being blurred.
Therefore, comparing FW-Net and FW-Net without \(\mathcal{L}_3\) loss, it makes the edge and structure information better integrated in fused image.
% \(\mathcal{L}_3\) loss can not only retain the semantics of the source images, but also make the fused image boundary not be blurred.
% Therefore, compared with FW-Net and FW-Net without \(\mathcal{L}_3\) loss, it can improve MI, AB and SSIM.

\section{Conclusion}
% Medical images play an important role in clinical diagnosis, and it is necessary to analyze medical images of different modalities.
% However, existing medical image fusion approaches ignore the semantics of medical images, which makes the fused images incomprehensible.
% In this paper, we propose a semantic-based approach to fuse medical images: extract the semantics of source images of different modalities, map semantics to a new semantic space, and fuse images in the new semantic space.
In this paper, we propose a new evaluation index to measure the semantic loss of the fused image, and a novel framework which combines semantic loss and structural loss is used for medical image fusion.
We perform a detailed quantitative and qualitative evaluation as well as ablation study.
In contrast to five state-of-the-art approaches, our approach effectively solves semantic conflicts and produces visually pleasing images.
Our approach is promising and expected to be applied in the clinical setting in future.

\bibliographystyle{IEEEbib}
\bibliography{icme2020template}

\end{document}